\documentclass{vgtc}                          

\ifpdf
  \pdfoutput=1\relax                   
  \usepackage{graphicx}                
  \DeclareGraphicsExtensions{.pdf,.png,.jpg,.jpeg} 
\else
  \usepackage{graphicx}                
  \DeclareGraphicsExtensions{.eps}     
\fi%

\graphicspath{{figures/}{pictures/}{images/}{./}} 

\usepackage{color,soul}
\usepackage{microtype}                 
\PassOptionsToPackage{warn}{textcomp}  
\usepackage{textcomp}                  
\usepackage{mathptmx}                  
\usepackage{times}                     
\usepackage{cite}                      
\usepackage{tabu}                      
\usepackage{booktabs}                  

\usepackage{xspace}
\usepackage{amsmath}
\usepackage{chngcntr}
\newcommand{\toolname}{RMExplorer\xspace}
\newcommand{\chest}{C\textsubscript{2}HEST\xspace}
\onlineid{}

\vgtccategory{Research}

\vgtcinsertpkg

\title{\toolname: A Visual Analytics Approach to Explore the Performance and the Fairness of Disease Risk Models on Population Subgroups}

\author{Bum Chul Kwon\thanks{e-mail: bumchul.kwon@us.ibm.com}\\ %
        \scriptsize IBM Research %
\and Uri Kartoun\\ %
     \scriptsize IBM Research %
\and Shaan Khurshid\\ %
     \scriptsize \centering Broad Institute
\and Mikhail Yurochkin\\ %
     \scriptsize IBM Research %
\and Subha Maity\\ %
     \scriptsize University of Michigan %
\and Deanna G Brockman\\ %
     \scriptsize \centering Broad Institute
\and Amit V Khera\\ %
     \scriptsize \centering Broad Institute
\and Patrick T Ellinor\\ %
     \scriptsize \centering Broad Institute
\and Steven A Lubitz\\ %
     \scriptsize \centering Broad Institute
\and Kenney Ng\thanks{e-mail: kenney.ng@us.ibm.com}\\ %
     \scriptsize IBM Research %
}

\teaser{
  \centering
  \frame{
  \includegraphics[trim={0 0.1cm 0 0.1cm},clip,width=\linewidth]{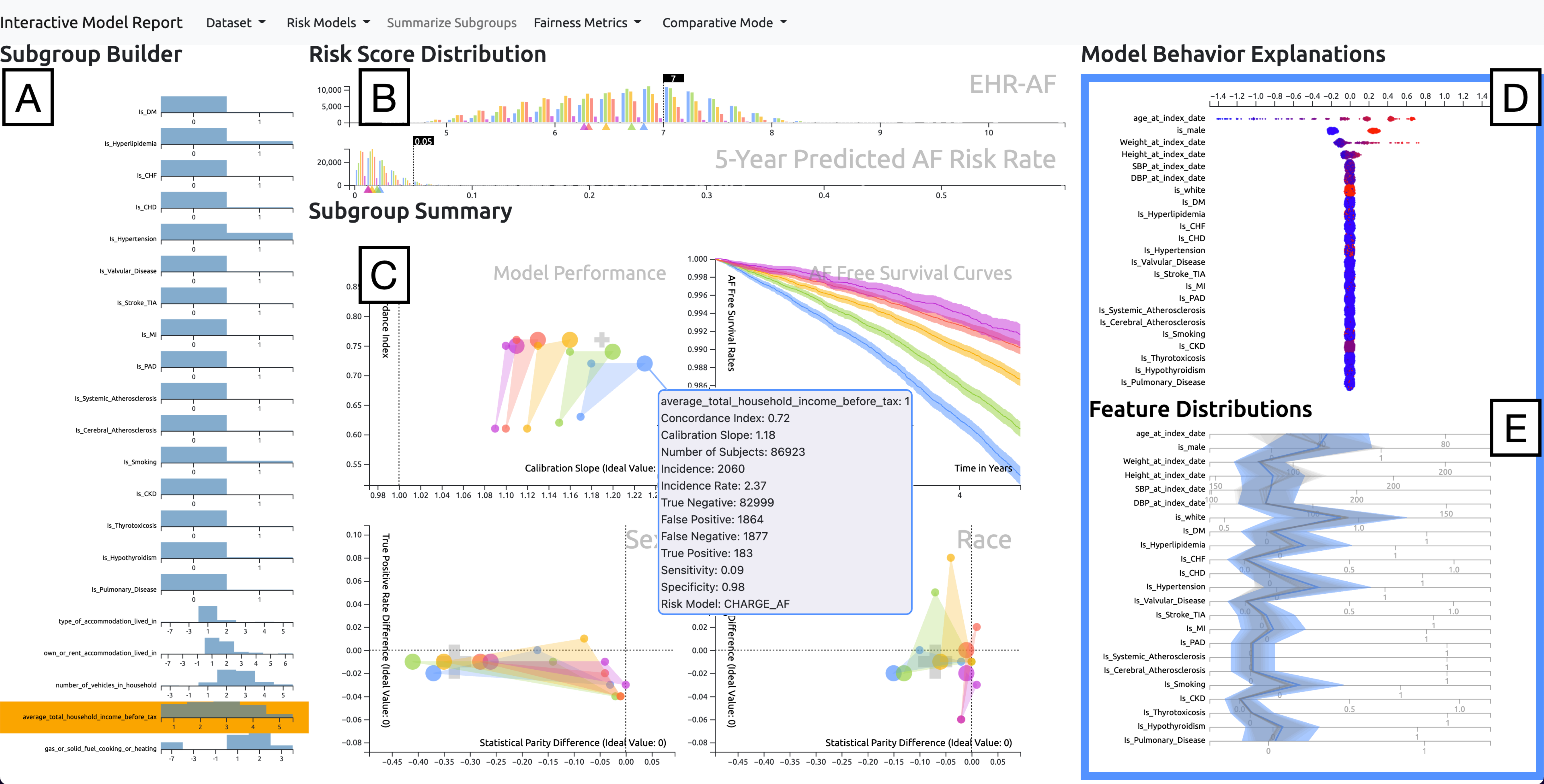}
  }
  \caption{\toolname: (A) Subgroup builder shows the distribution of patients by variables that can be used to create subgroups; (B) Risk score distribution displays the summary of risk scores and the estimated disease onset rate; (C) Subgroup summary presents the model performance (top) and the fairness of subgroups (bottom); (D) Model behavior explanations provide feature contributions to the risk scores; (E) Feature distributions show the mean and confidence intervals of a selected subgroup across variables.}
  \label{fig:overview}
}

\abstract{
Disease risk models can identify high-risk patients and help clinicians provide more personalized care. However, risk models developed on one dataset may not generalize across diverse subpopulations of patients in different datasets and may have unexpected performance. It is challenging for clinical researchers to inspect risk models across different subgroups without any tools. Therefore, we developed an interactive visualization system called \toolname (Risk Model Explorer) to enable interactive risk model assessment. Specifically, the system allows users to define subgroups of patients by selecting clinical, demographic, or other characteristics, to explore the performance and fairness of risk models on the subgroups, and to understand the feature contributions to risk scores. To demonstrate the usefulness of the tool, we conduct a case study, where we use \toolname to explore three atrial fibrillation risk models by applying them to the UK Biobank dataset of 445,329 individuals. \toolname can help researchers to evaluate the performance and biases of risk models on subpopulations of interest in their data.
} 


\CCScatlist{
  \CCScatTwelve{Human-centered computing}{Visu\-al\-iza\-tion}{}{}
}

\begin{document}


\firstsection{Introduction}

\maketitle

Disease risk models trained within a specific dataset may not generalize well to new, unseen datasets in which patient characteristics may differ - including observation periods, demographics, and healthcare regulations/policies - from the training dataset. 
In addition, the model may learn and propagate inherent, systematic biases captured in the training data, thus amplifying fairness issues. 
For instance, machine learning models trained with data at a university hospital, which often includes a majority of patients with higher income, younger age, and white race, may not be generalizable to more diverse populations~\cite{gianfrancesco2018potential}. 
Medical decisions that are based on biased models may lead to worse outcomes disproportionately affecting specific racial, gender, and socioeconomic groups~\cite{mehrabi2021survey}. 
Therefore, it is important for clinicians to understand the machine learning models in terms of both performance and fairness. 
Furthermore, because patient populations can be very heterogeneous, it is also important to understand how the model behaves for different subpopulations.

Recently, several approaches were proposed to summarize the fairness of model outcomes for given attributes. For instance, the AI Fairness 360 Toolkit~\cite{bellamy2018ai} is an open-source framework that includes various model-agnostic fairness metrics. 
However, it is computationally challenging to generate performance and fairness measures for all possible combinations of subgroups. 
Furthermore, it is cognitively demanding for users to analyze the results. 
Therefore, it is important to develop tools to help clinical investigators build subgroups and evaluate results easily. 
Recently, researchers investigated the concept of a model report card which summarizes how a model was trained and how it performs on a fixed set of subgroups in a static report~\cite{mitchell_model_2019, arnold_factsheets_2019}. 
These approaches are significant milestones, yet they do not allow users to dynamically apply the model to their own datasets, define custom subgroups, and interactively explore how the model behaves across the subgroups.

Visual analytics approaches are recognized as a potential solution to explore large-scale data in healthcare~\cite{caban_visual_2015, west_innovative_2015}. 
Visual analytics, which combines computational methods with interactive visualizations, can be used to generate insights and knowledge from data and to enhance trust~\cite{sacha2014knowledge, sacha2016role, chatzimparmpas_state_2020, liu2017towards}.
Interactive visualizations powered by machine learning algorithms on longitudinal health records have demonstrated usefulness for understanding disease subtypes~\cite{kwon2018clustervision}, analyzing heterogeneous disease progression patterns~\cite{kwon2020dpvis}, and predicting health outcomes~\cite{kwon2019retainvis}.
Recently, visual analytics applications are developed to explore fairness in ranking decision~\cite{ahn_fairsight_2020}, image classification~\cite{kwon_dash_2022}, and others.
However, previous studies have not extensively studied visual analytics approaches to investigate the performance and fairness of risk models on healthcare data.

To address the important, unmet need, we propose a visual analytics approach called \toolname (Risk Model Explorer) to facilitate the exploration of performance heterogeneity and fairness. The system includes interactive visualizations combined with various computational approaches. The system can be applied to explore the performance and fairness of any risk models and any appropriate datasets. To demonstrate the applicability and usefulness of the system, we provide a use case on a large prospective cohort dataset (UK BioBank~\cite{sudlow_uk_2015}) with validated atrial fibrillation (AF) risk models, namely, EHR-AF~\cite{khurshid_performance_2021}, CHARGE-AF~\cite{alonso_simple_2013}, and \chest~\cite{li_simple_2019}.

\section{Visual Analytics System Design}

\toolname allows clinical researchers to apply existing risk models to their patient datasets, to visually explore the performance and fairness of models within subgroups defined by users, and to probe why the model predicts different outcomes for each subgroup. 
In the following sections, we describe the design of \toolname.

\subsection{Users, Tasks, and Requirements}

All authors of the study in diverse disciplines, including  cardiology, machine learning, and visualization, characterized the tasks and requirements of disease risk model analysis.
The intended users, clinical researchers and data scientists, want to evaluate disease risk models.
First, they assess the performance of the model by computing risk scores and prediction accuracy. 
The risk scores are converted to estimated disease onset rates so that users can compare them with other previously reported onset rates.
Second, users inspect inherent biases of the scores with respect to various patient subpopulations that they define using various demographic and clinical variables.
Third, they also comparatively analyze different metrics for different risk models to test the robustness of the measures.
The process repeats iteratively until they gain insights into the risk models.

To support the analysis, the system requires interactive visualizations, where users can view the summary of various metrics about performance and fairness.
In particular, researchers need to define subgroups and compute various metrics on each of them so that they understand the risk models that are applied to the subpopulations.

\subsection{Subgroup Builder}

Subgroup Builder shows the distribution of patients over a specified set of variables, such as comorbidities, age, sex, and race. The initial view shows the overall summary of the entire population. Users then can select a set of variables which will be used to create subgroups. For instance, if users want to create a subgroup based on hyperlipidemia disease status in the health records, they can select the variable ‘is\_Hyperlipidemia’. If users select more than one variable, the system will generate all possible combinations of the selected variables, each of which becomes a mutually exclusive group of patients that satisfy that combination of conditions. Subgroups without any patients are automatically discarded. Figure~\ref{fig:overview}~(A) shows that the user chose one variable with five discrete levels, which subsequently generated five subgroups for analysis in other views.

\subsection{Risk Score Distribution}

\toolname requires users to provide risk models that are previously trained. 
The models are provided as functions that generate risk scores and probabilities of diseases during a follow-up time window (e.g., 5 years since the baseline). 
\toolname can then compute the metrics for selected patients using the model covariates on the fly.

Risk Score Distribution (Figure~\ref{fig:overview}~(B)) shows distributions of a chosen risk score (top) and its predicted risk rate at a future date (bottom) for the selected subpopulation. 
Each of the two charts provides a threshold handle that represents a constant value to convert the continuous risk score to the binary predicted label (high-risk vs low-risk). 
The two thresholds are synchronized: if users adjust one threshold then the other is adjusted accordingly.
As users update the threshold, the system recalculates the predicted labels, which are used by the performance and fairness metrics in Subgroup Summary. Users can at any point switch to other risk models, which resets the view with the chosen scores and predicted risk rates. Figure~\ref{fig:overview}~(B) shows the distribution of EHR-AF, a risk model that predicts the time to onset of atrial fibrilation since the baseline of covariates measured, and 5-year predicted disease risk rate.

\subsection{Subgroup Summary: Performance}

Subgroup Summary presents the performance (the two charts on top) and the fairness (the other two charts at the bottom) of the risk score model on the subgroups defined by users as Figure~\ref{fig:overview}~(C) shows. 

The risk model performance is shown in two different plots located on top. The first chart on the left initially shows a scatter plot of subgroups as points, colored according to subgroups, on two-dimensional space with calibration slope on the x-axis and concordance index on the y-axis. Concordance index, also known as Harrell’s C-index, is a measure that evaluates the discrimination performance of risk models in survival analysis, where data are censored (e.g., lost to follow up)~\cite{harrell1982evaluating}. Calibration slope is a measure of the relation between predicted risk and observed incidence, where a value of one indicates optimal calibration. Among models that are reasonably calibrated-in-the-large, values higher than 1 indicate underestimation while values lower than 1 indicate overestimation of the actual risk~\cite{van2016calibration}. With these two measures, users can gain insights into how well the model discriminates disease risk in each subgroup and how well the model estimates are calibrated within each subgroup. Users can expand the view by showing other risk models in the view, which creates polygons, colored according to subgroups (Figure~\ref{fig:overview}~(C)). The vertices of each polygon represent the respective risk models. By switching to the comparative mode, users can investigate the similarities and differences of multiple risk models across and within subgroups.

The second chart on the right shows a disease-free survival plot stratified by subgroups. The plot is generated by fitting the patients’ disease-free survival data using the Kaplan-Meier method. By viewing the chart, users can understand the relative differences in estimated risks over time among different subgroups.

\subsection{Subgroup Summary: Fairness}
The two charts at the bottom show fairness metrics for subgroups that users have chosen. Our system supports group and individual fairness metrics, and users can choose any two at a time.
As Figure \ref{fig:overview}~(C) shows, the system provides two group fairness metrics: statistical parity difference on the x-axis and true positive rate difference on the y-axis computed using the AI Fairness 360 Toolkit~\cite{bellamy2018ai}. 
For group fairness metrics, each of the charts presents the fairness in terms of a protected attribute. 
A protected attribute (e.g., gender, race, socioeconomic level) is an application-specific, user-selected attribute of patients across the categories of which it is desirable to achieve equality in terms of benefit received.
A value of 0 for these measures indicates an equal performance of the model for any given protected attribute.
Let’s say we define race as a target protected attribute, then statistical parity difference computes the difference between the proportion of the predicted patients diagnosed (by the risk model) with the target disease for patients with nonwhite race and that for patients with white race within each subgroup. 
True positive rate difference measures the difference in the accuracy of risk models that predict the disease of patients between the nonwhite and the white race groups within each subgroup.
For individual fairness, we compute the individual fairness violation rate, which provides the proportion of individuals whose outcome predictions are sensitive to the protected attributes~\cite{maity2021statistical, yurochkin2019training}.
By viewing the fairness charts, users can understand the differences between subgroups in terms of key fairness metrics for each protected attribute.

Similar to the first chart of the model performance, the fairness plot initially shows a scatterplot of points, colored by subgroups, over the two dimensions of selected fairness measures. Users can switch to the comparative mode so that the view shows polygons, colored by subgroups, over the two dimensions.

\subsection{Model Behavior Explanations}

Model Behavior Explanations aim to show why the model behaves the way it does for a selected subgroup. The view opens when users select a subgroup in the Subgroup Summary. The view includes two visualizations: the SHAP (SHapley Additive exPlanations) beeswarm plot and the feature distribution chart. 

SHAP is a model-agnostic interpretability approach that generates a feature importance value, called a SHAP value, for each prediction~\cite{lundberg2017unified}. 
Once the tool generates the SHAP values for all predictions made for the patients in the selected subgroup, it plots them using the beeswarm plot. 
The plot shows dots (patients) per each row (a feature). 
The color of each dot indicates the normalized feature value, ranging from minimum (blue) to maximum (red) for the corresponding feature. 
The horizontal position of each dot represents the SHAP value. 
The vertical position is jittered to show the distribution of points along the x-axis without occlusion. 
To avoid the visual clutter from overplotting and to speed up computation, the tool randomly samples a fraction (i.e., 10\%) of the subgroup population to generate the beeswarm plot. 
Figure~\ref{fig:overview}~(D) shows the relationship between SHAP values and feature values of 23 variables.

To summarize the feature distributions of the selected subgroups, we use a variation of parallel coordinates, called Parallel Trends, which shows a central measure (e.g., mean) and a dispersion (e.g., standard deviation) of a group of instances with band crossing axes~\cite{kwon2018clustervision}. 
The view highlights the selected subgroup while displaying other subgroups in a grey color. Using the visualizations, users can compare the feature distributions across multiple subgroups. 
Figure~\ref{fig:overview}~(E) shows the parallel trends of the blue subgroup in the context of other subgroups in the background.

\section{Case Study: Atrial Fibrillation Risk Models}

In this section, we provide a case study on risk models of atrial fibrillation (AF), a common cardiovascular condition that is a leading cause of stroke and heart failure in the elderly. 
The case study was performed by all authors, where the leading author mainly operated the tool, and all authors, including the five domain experts who are actively involved in clinical research in cardiac disease risk models, analyzed the results using visualizations.
In this case study, we used three representative AF risk score models: EHR-AF~\cite{khurshid_performance_2021}, CHARGE-AF~\cite{alonso_simple_2013}, and \chest~\cite{li_simple_2019}, each of which provides scores for individual patients and the 5-year risk of AF using their respective covariates, such as demographics (e.g., age, race, sex) and AF-risk factors (e.g., hypertension, hyperlipidemia, heart failure).

For the dataset, we extracted a patient cohort from the UK Biobank (research application 50658), an observational study that enrolled over 500,000 individuals between 2006 and 2010~\cite{sudlow_uk_2015}. 
To be consistent with the original design of CHARGE-AF, we excluded individuals under age 45 at enrollment, individuals with prevalent AF, and individuals with missing data for height, weight, systolic blood pressure (SBP), and diastolic blood pressure (DBP) at baseline (Table~\ref{tab:cohortselection}). 
The final cohort contains 445,329 patients with 7,407 (1.66\%) having incident AF within 5 years after their enrollment. 

We hypothesized that the risk model may estimate AF risk disproportionately with respect to an individual patient’s socioeconomic status because there may be association between disease incidence and household income~\cite{larosa_association_2020, misialek_socioeconomic_2014, guhl_association_2019}.  
The variable includes five income ranges as discrete integer levels: 1) less than £18,000; 2) £18,000 to £30,999; 3) £31,000 to £51,999; 4) £52,000 to £100,000; 5) Greater than £100,000. 
We discarded those who did not report their income by selecting ``Do not know'' and ``Prefer not to answer''.

\begin{figure}[t]
    \centering
    \includegraphics[width=\columnwidth]{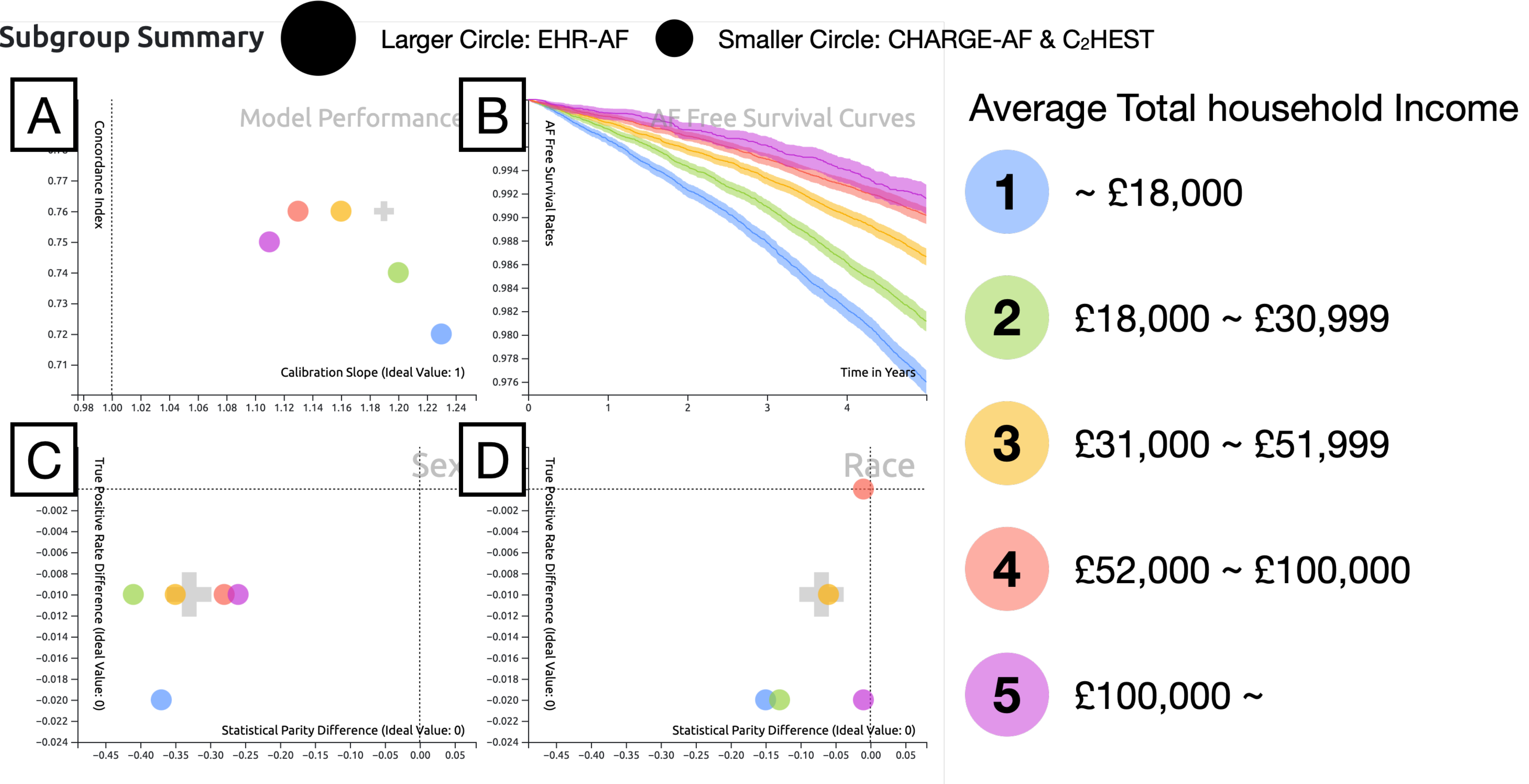}
    \vspace{-.7cm}
    \caption{The subgroup summary shows differences in model performance and fairness among the five income subgroups: (a) model performance; (b) atrial fibrillation (AF) free survival curves; (c) fairness on sex; (d) fairness on race.}
    \label{fig:subgroup_summary_default}
    \vspace{-.7cm}
\end{figure}

A user launches \toolname to explore the three AF risk models. Using Subgroup Builder, they selected the variable called ``Annual Total Household Income Before Tax''. 
We select EHR-AF as the risk model to begin the analysis.
As the user clicks on ``Summarize Subgroups'', Subgroup Summary provides four visualizations of the five subgroups as Figure~\ref{fig:subgroup_summary_default} shows. 
First, the model performance view shows that concordance index is lower for patients in lower income groups (blue and green) and generally higher for those with higher income (Figure~\ref{fig:subgroup_summary_default}~(A)). 
Similarly, the calibration slope for the higher income group is closer to the ideal value (the dotted, vertical line), 1, but the lower income group is more distant from 1.
In other words, users can hypothesize that the EHR-AF risk model makes more accurate decisions for higher income groups.
In addition, the AF free survival curves in Figure~\ref{fig:subgroup_summary_default}~(B) show that the higher income groups have longer AF-free survival than the others. 

The fairness plots summarize the fairness metrics of the models for the income subgroups. 
The fairness plot on Sex (Figure~\ref{fig:subgroup_summary_default}~(C)) shows that all subgroups have negative values for both true positive rate differences and statistical parity differences. 
This indicates that the true positive rates for females are lower than those for males, and the proportion of predicted AF for females is lower than males for all subgroups. 
Similarly, the fairness plot on Race (Figure~\ref{fig:subgroup_summary_default}~(D)) shows that subgroups tend to have negative values for both metrics as well. 
Interestingly, the lower income groups show greater disparities in the statistical parity difference between white and non-white patients than the higher income groups.

\begin{figure}[t]
    \centering
    \includegraphics[width=\columnwidth]{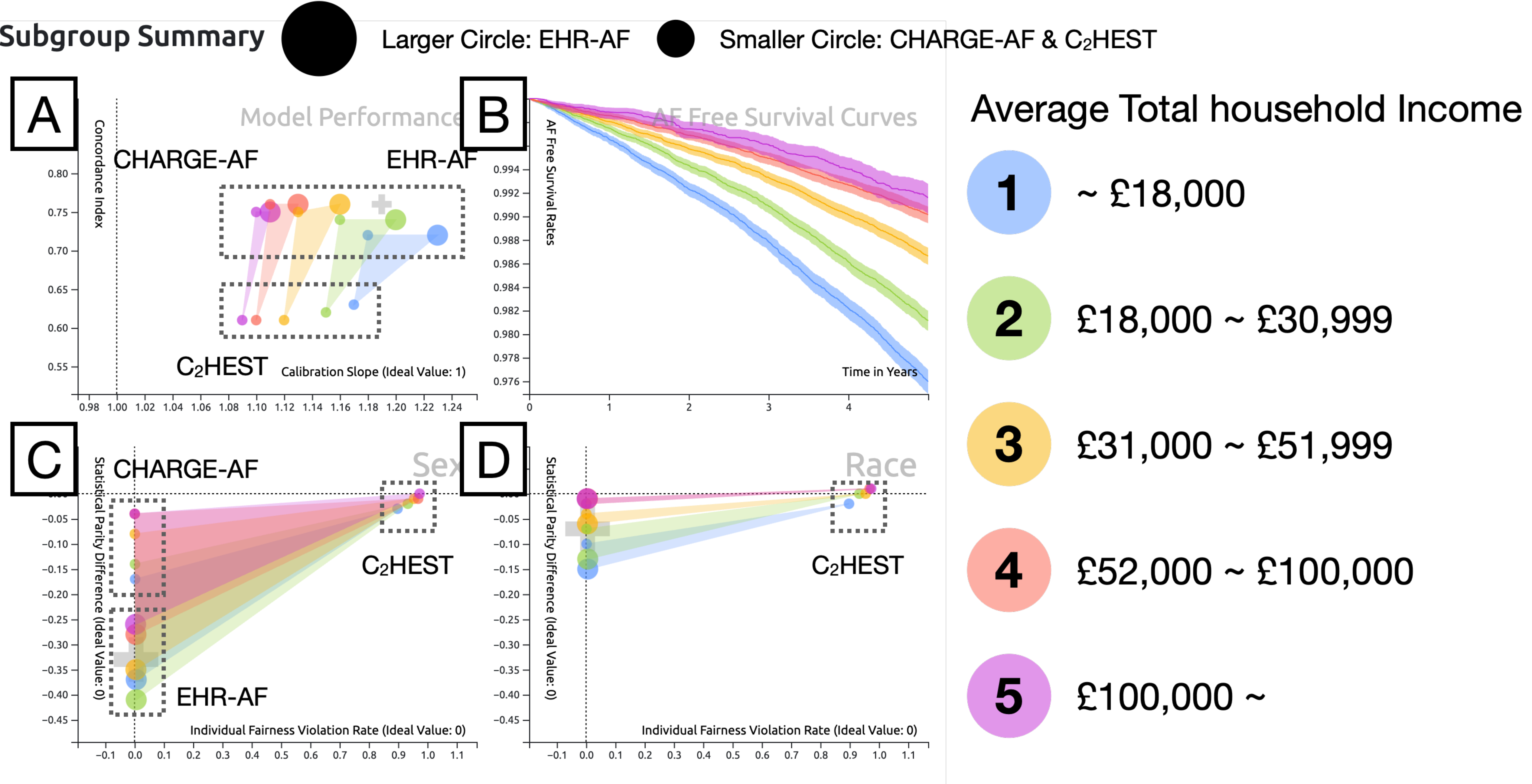}
    \vspace{-.8cm}
    \caption{The subgroup summary compares EHR-AF, CHARGE-AF, and \chest. The polygon areas with distinct colors show differences in the scores across subgroups, and the coordinate location of vertices per each polygon summarizes the differences in scores among risk models within each subgroup.}
    \label{fig:subgroup_comparison}
    \vspace{-.75cm}
\end{figure}

What about the two other risk models? 
The user switches to the comparative mode to explore the differences among risk models. 
Figure~\ref{fig:subgroup_comparison}~(A) illustrates that the five polygons are shown next to each other, starting from low income groups (right) to high income groups (left). 
It also shows that each triangle has its vertices in relatively consistent positions, as CHARGE-AF on top-left, EHR-AF on top-right, and \chest at the bottom. 
It shows that \chest has lower performance than the other risk models consistently across subgroups. 
As Figure~\ref{fig:subgroup_comparison}~(C) shows, the fairness scores of the risk models are separated in their relative positions within each subgroup with CHARGE-AF on top left, EHR-AF on bottom left, and \chest on top right. 
\chest shows very high values for individual fairness violation rates, close to 0.9, compared to other risk scores, close to 0. 
This indicates that \chest is likely to change predictions for individuals if the input values are perturbed with respect to sex. 
The similar insight can be drawn for the race attribute in Figure~\ref{fig:subgroup_comparison}~(D).

The user clicks through subgroups to explore their behaviors in the SHAP plot and the feature summary. 
The SHAP plots show that the age of patients is the most influential factor in risk estimation across the income subgroups.
By their distributions, they can see that the first four variables, age, sex, weight, and height, have non-zero values of SHAP, which indicates their relative importance to the EHR-AF score. 
As the plot shows, the estimated risk is higher for older patients. 
The feature summary plots show that age is indeed the variable that differentiates the income subgroups, as lower income groups have relatively higher mean age. 
The feature distribution shows that height is lower for lower income groups (1,2) than other income groups.
With the consistent mean weight across subgroups, the user speculates that lower income groups might have higher body mass indices (BMIs), which may lead to higher AF risk scores.

\begin{figure}[t]
    \centering
    \includegraphics[width=\columnwidth]{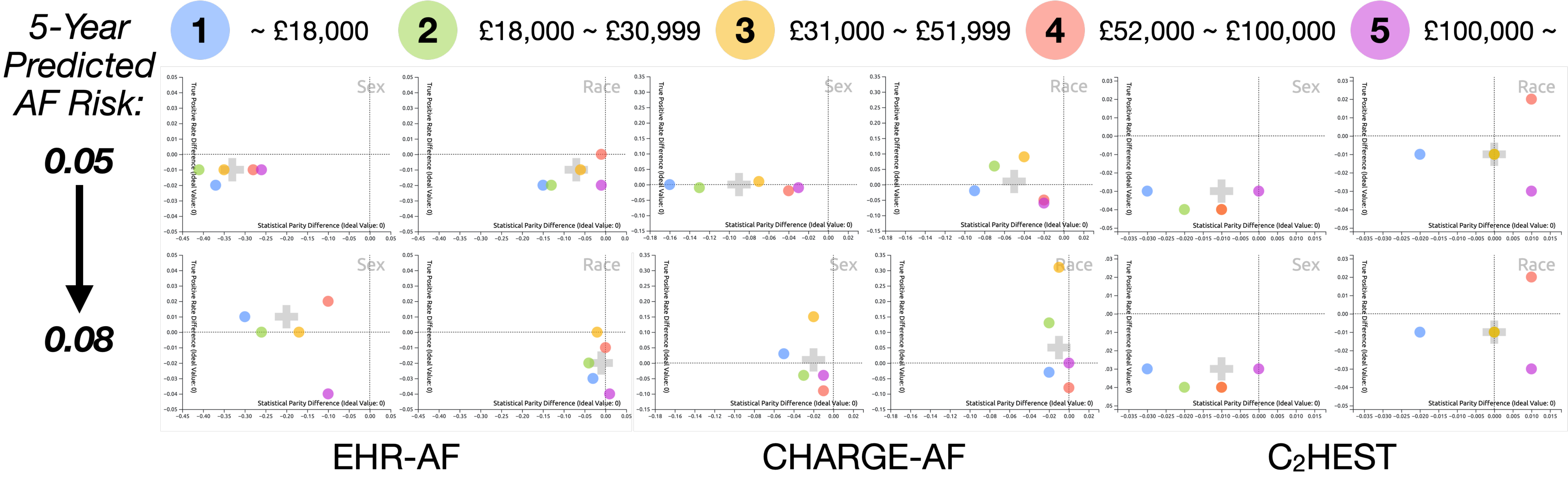}
    \vspace{-.8cm}
    \caption{The fairness plots show differences when the 5-year predicted risk threshold is adjusted from 0.05 to 0.08.}
    \label{fig:af_risk_adjustment}
    \vspace{-.75cm}
\end{figure}

To explore how fairness measures vary as a function of 5-year predicted AF risk thresholds and income levels, the user decides to experiment with multiple threshold values on Risk Score Distribution and observe their results (Figure~\ref{fig:af_risk_adjustment}). 
By default, the system shows summary results with 0.05 as the 5-year risk score threshold because it is a standard threshold used in previous studies~\cite{khurshid_performance_2021}. 
The user increased the value to 0.08. 
The updated fairness plots (Figure~\ref{fig:af_risk_adjustment} bottom) show some differences in the three different risk models among the five income-based subgroups after the threshold adjustment has been made. 
In EHR-AF, the true positive rate differences for females are now higher than or equal to those for males in the income subgroups 1, 2, 3, and 4. 
This means that the model with the updated threshold makes decisions for female patients more accurately than or at a similar level to male patients in those subgroups. 
The fairness plot on Race shows the gap that previously existed on statistical parity (x-axis) is now narrower than before as all subgroups are closer to 0. 
On the other hand, in CHARGE-AF, the updated threshold vertically widened the spread of subgroups with respect to true positive rate differences in sex and race. 
However, it also narrowed the spread in statistical parity difference. 
For \chest, the user does not find any differences before and after the updated thresholds for both fairness measures with respect to sex or race. 
The experiment shows that the updated threshold can affect the fairness metrics differently for different risk models across the income subgroups.

\section{Discussion and Future Work}

This study presents \toolname that allows users to define subgroups, assess risk model performance and fairness based on the subgroups, and understand the model behavior with feature distributions and feature importance measures. 
The multiple, coordinated visualizations can help users gain insights into how the model performs differently for different subgroups. 
The case study on atrial fibrillation risk models demonstrates the usefulness of the interactive visualizations combined with various computational approaches for generating clinically relevant findings. 
The interactive visualizations provide some design implications for tools that support subgroup creation and exploration with respect to risk models. Subgroups are the keys to examining the risk models, and visualizations need to provide users with abilities to create, modify, and compare subgroups by various means. In this study, we provided the subgroup builder, where users can create customizable mutually exclusive subgroups. Future work can investigate various methods to computationally discover and recommend subgroups for further analysis. Interactive visualizations need to summarize the subgroups using patient characteristics and allow users to modify them if necessary. In addition, we present polygons in scatter plots to enable comparison of multiple risk models. The approach can be revisited and improved for a greater number of subgroups.

The visualizations show the performance and fairness of subgroups using various metrics. Our work provides an example using three risk score models for survival analysis, which are commonly used in clinical research. The survival curve is a specific visualization that is tailored for this type of analysis. However, the same visualization strategy can be applied to other prediction models with different performance metrics (e.g., AUC score). The fairness plots are designed to show disparity of subgroups in terms of two group fairness and an individual fairness metrics. Future work can investigate the usefulness of other fairness metrics. 

The SHAP plot and the feature distribution summary provide explanations as to why the model behaves in certain ways for different subgroups. There are various other explanation approaches (e.g., LIME~\cite{ribeiro2016should}), which can be added to the visualization pipeline with additional coding. In addition to the model-agnostic techniques, one may derive some explanations directly from the model when it uses the attention mechanism for deep learning models \cite{hoover2019exbert}. Then, the visualizations need to be adapted to show the attention scores for relevant input features instead. Future work can investigate various ways to combine the attention-based model outputs with other important metrics for the subgroup analysis. Future work can also evaluate \toolname against other visual analytics systems, such as FairSight~\cite{ahn_fairsight_2020}, for model performance and fairness inspection tasks.

\acknowledgments{
We are indebted to the UK Biobank and its participants who provided biological samples and data for this analysis. UK Biobank analyses were conducted via applications 7089 and 50658.
}

\bibliographystyle{abbrv-doi}

\bibliography{rme}

\clearpage

\appendix
\counterwithin{figure}{section}
\counterwithin{table}{section}

\section{Appendix}

\begin{figure}[h]
    \centering
    \includegraphics[width=\columnwidth]{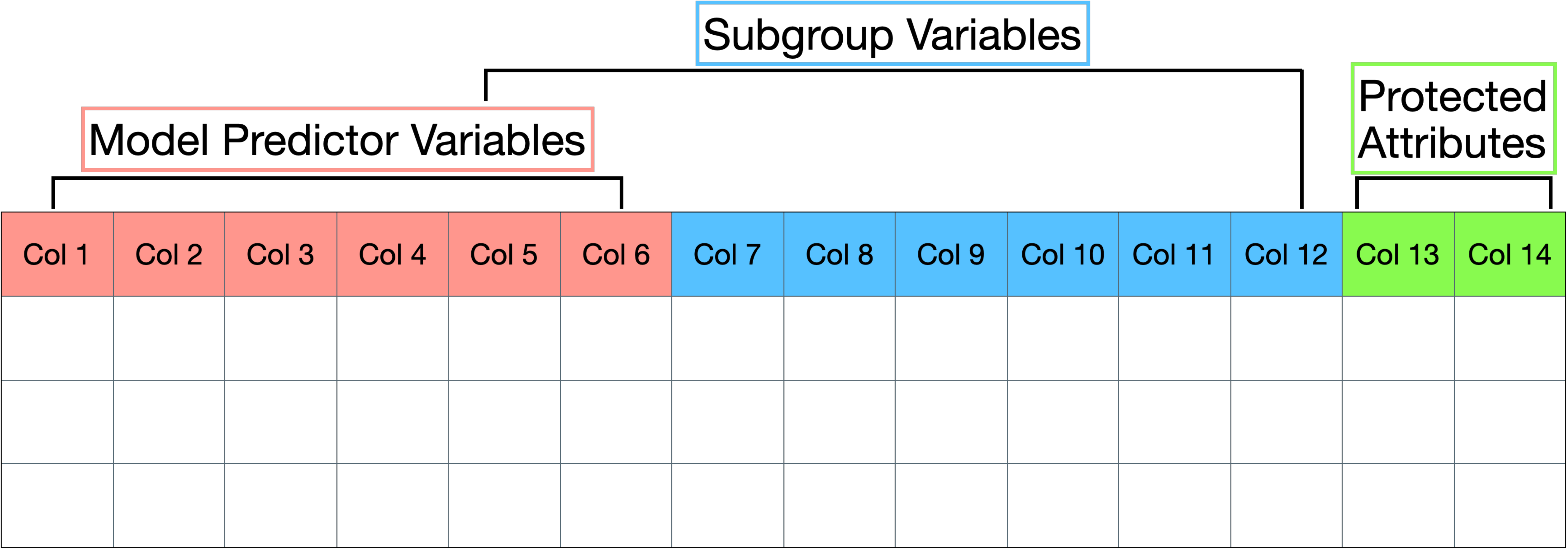}
    \caption{A diagram illustrating the description of datasets by their columns. The columns can be categorized into the three types: i) predictor variables; ii) subgroup variables; iii) protected attributes.}
    \label{fig:model_desc}
\end{figure}

\begin{figure}[h]
    \centering
    \frame{\includegraphics[width=\columnwidth]{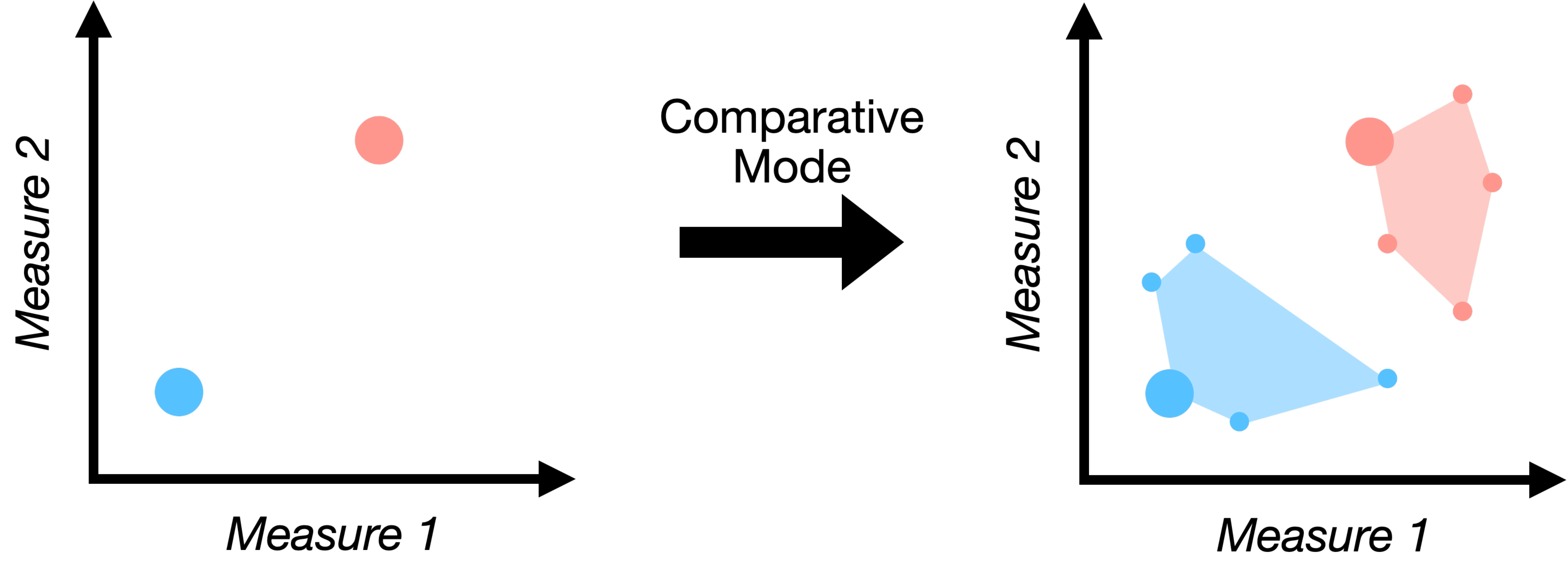}}
    \caption{Illustration of the comparative mode of scatter plots used for the performance and fairness of subgroups in Subgroup Summary. Initially, each scatter plot shows dots colored by subgroups on two continuous measures. As users switch to the comparative mode, the chart introduces other risk models to the chart by drawing a polygon per subgroup (color). Each vertex of the polygon represents a different risk model.}
    \label{fig:comparative_mode}
\end{figure}

\begin{figure}[t]
    \centering
    \includegraphics[width=\columnwidth]{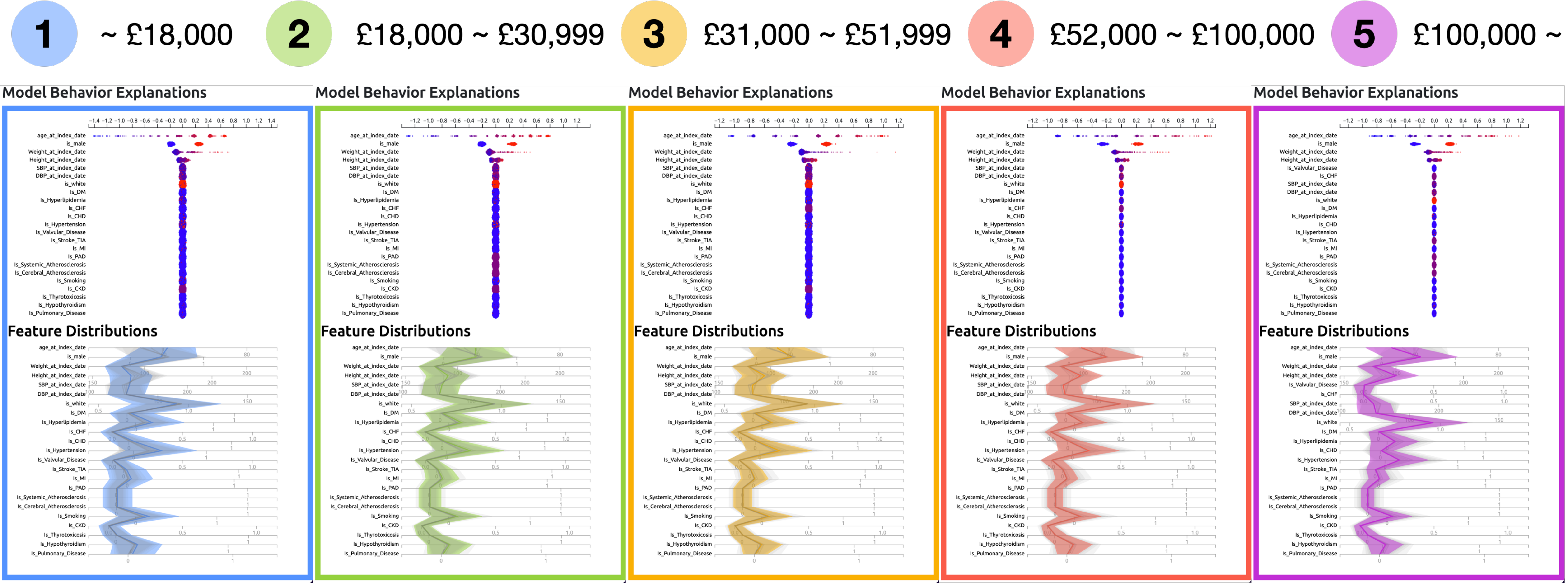}
    \caption{The model behavior explanation plots, the SHAP plot and the feature distribution summary, for the five income subgroups from low (left) to high (right). The feature distributions show some differences in age of patients at the baseline (dotted areas).}
    \label{fig:behavior_explanation}
\end{figure}

\begin{table}[h]
\small
\centering
\caption{Population selection from the UK Biobank (incident AF).}
\label{tab:cohortselection}
\begin{tabular}{@{}cl@{}}
\toprule
Number of patients & Selection criteria                                         \\ \midrule
502,521            & All registered patients.                                   \\
456,793            & Age $\geq$ 45 at baseline (``date of enrollment'').               \\
453,573            & Complete values: height, weight, SBP, and DBP at baseline. \\ 
445,357            & No history of AF on or prior to baseline.                  \\
445,329            & Exclude patients who asked to opt out from the study.      \\ \bottomrule
\end{tabular}
\end{table}

\begin{table}[h]
\small
\centering
\caption{Patient population characteristics of the study cohort.}
\label{tab:populationcharacteristics}
\begin{tabular}{@{}lc@{}}
\toprule
Characteristics & $\%$ or Mean (Standard Deviation)                                        \\ \midrule
Patient Demographics at baseline                                \\ \midrule
Female (\%) &	55.0  \\
Age (years)	& 58.4 (7.0)  \\
White race (\%)	& 94.7  \\
Smoking (\%) &	10.7  \\
SBP (mmHg) &	138.9 (18.6)  \\
DBP (mmHg) &	82.5 (10.1)  \\
Height (cm)	& 168.2 (9.2)  \\
Weight (kg)	& 77.9 (15.8)  \\
Hypertension (\%) & 	30.5  \\
Diabetes (\%) &	2.5  \\
Hyperlipidemia (\%)	& 15.7  \\
Heart failure (\%)	& 0.4  \\ \midrule
Annual Income Level (\%): \\ 
\hspace{3mm}1. Less than £18,000 &	19.5 \\
\hspace{3mm}2. £18,000 to £30,999 &	22.1 \\
\hspace{3mm}3. £31,000 to £51,999 &	21.9 \\
\hspace{3mm}4. £52,000 to £100,000 &	16.9 \\
\hspace{3mm}5. Greater than £100,000 &	4.5 \\
\hspace{3mm}Do not know or prefer not to answer &	15.1 \\ \bottomrule
\end{tabular}
\end{table}

\begin{table}[ht]
\small
\centering
\caption{Multivariate model coefficients (row) for the three scores (columns): EHR-AF, CHARGE-AF, and \chest. Each risk model comprises different combinations of predictor variables. \textnormal{-} indicates that the corresponding variable (row) was not used to predict the corresponding score (column).}
\label{tab:model_coef}
\begin{tabular}{@{}lccc@{}}
\toprule
Variables & EHR-AF & CHARGE-AF & \chest                     \\ \midrule
Male &	0.137 & - & - \\
Age, per 10-yr increase	& 1.494 & 1.016 & - \\
Age $\geq$ 75 & - & - & 2 \\
Squared age, per 10-yr increase	& -0.048  & - & - \\
White &	-0.208 & 0.465 & -  \\
Smoking &	0.152 & 0.359 & - \\
Height, per 10-cm increase & -0.231 & 0.248 & - \\
Squared height, per 10-cm increase	& 0.012 & - & - \\
Weight, per 15-kg increase	& -0.050 & 0.115 & - \\
Squared weight, per 15-kg increase & 0.021 & - & - \\
Systolic BP, per 20 mm Hg increase & - & 0.197 & - \\
Diastolic BP, per 10 mm Hg increase & - & -0.101 & - \\
Diastolic blood pressure $\geq$ 80 mm Hg & -0.104 & - & - \\
Diabetes & - & 0.237 & - \\
Myocardial infarction & - & 0.496 & - \\
Hypertension	& 0.106  & - & 1 \\
Hyperlipidemia & -0.156  & - & - \\
Heart failure & 0.563 & 0.701 & 2\\
Coronary heart disease & 0.210 & 0.349 & 1 \\
Valvular disease & 0.487 & - & - \\
Previous stroke/TIA & 0.132 & - & - \\
Peripheral artery disease & 0.126  & - & -\\
Pulmonary disease & - & - & 1 \\
Chronic kidney disease & 0.279  & - & - \\
Hypothyroidism & -0.138  & - & 1 \\ \midrule
5-year estimated risk variables: \\ 
\hspace{5mm}c &	 0.971 & 0.972 & 0.975 \\
\hspace{5mm}bias &	6.454 & 12.582 & 0.370 \\ \bottomrule
\end{tabular}
\end{table}

\subsection{Individual Fairness}

For individual fairness, we adopted the auditing procedure of \cite{maity2021statistical}. This procedure aims to find individuals that are similar to those in the original dataset while experiencing different model outcomes. Different decisions on a pair of similar individuals constitute an individual fairness violation. We report a fraction of data points where individual fairness is violated, i.e. the auditing algorithm \cite{maity2021statistical} manages to find similar individuals with predicted risk scores on the opposite sides of the user-specified threshold. The similarity of individuals is quantified using the fair distance learned from the data following the sensitive subspace approach described in Appendix B.1 of \cite{yurochkin2019training}. Their method measures the similarity of two individuals with covariates $x$ and $x'$ with a fair distance $d(x, x')$ which ignores all differences in a sensitive subspace and only accounts for the differences in the directions orthogonal to this subspace. The subspace consists of the directions which are most informative for the sensitive attributes and are obtained from logistic regression coefficients which predict a sensitive attribute from the other covariates. This fair distance treats two individuals  to be similar if they are similar in all respect, except for the sensitive attributes. For an individual with covariates $x$, the auditing method \cite{maity2021statistical} aims to find a similar individual $x'$ which is close to $x$ in terms of the fair distance and maximizes the prediction loss. Let $p(x)$ denote the predicted risk score for an individual ($x$), and $y(x)$ to be the indicator of whether an individual belongs to a high-risk group. Then $x'$ is obtained as
\[x' = 
\begin{cases}
\arg\min_{u} p(u) + \lambda  d(x, u) & \text{if } y(x) =1 , \\
\arg\max_u p(u) -  \lambda  d(x,u) & \text{if } y(x) = 0,
\end{cases}
 \]
and, the metric is calculated as \[
\text{individual fairness violation rate} = \frac{\text{\# individuals with } y(x) \neq y(x')}{\text{sample size}}\,.
\]

\subsection{Calculation of Cardiovascular Risk Models}

Each score is calculated by a sum of the product of the coefficient and value across the $n$ covariates for each patient:
\begin{equation}
    Score = \sum_{k=1}^{n}\beta_{k}x_{k}
\end{equation}
where $\beta$ is the coefficient for predictor covariate $x$. Then, using the baseline hazard and mean covariate estimates, we can compute 5-year estimated risk:
\begin{equation}
    5\textnormal{-}Year\:Estimated\:Risk = 1 - c^{\exp{(Score - bias)}}
\end{equation}
where c is a constant, representing average AF-free survival probability at 5 years.

The predictor variable coefficients for the three risk score models are shown in  Table~\ref{tab:model_coef}. To learn more about the model details, readers are advised to read the respective papers for EHR-AF~\cite{khurshid_performance_2021}, CHARGE-AF~\cite{alonso_simple_2013}, and \chest~\cite{li_simple_2019}.

\end{document}